\documentclass[pre,twocolumn,showpacs,amsmath,amssymb]{revtex4}
\def\aa{\AA\,} 
\def\eR{\textbf{\em R}\,}
\newcommand{\be}{\begin{equation}} 
\newcommand{\ee}{\end{equation}} 
\newcommand{\bea}{\begin{eqnarray}} 
\newcommand{\eea}{\end{eqnarray}}

\def\Journal#1#2#3#4{{#1} {\bf #2,} { #3} (#4)} 
 
\def\EPL{ Europhys. Lett.}

\def\JPCB{{ J. Phys. Chem.}B}

\def\JAP{ J. Appl. Phys.}

\def\PRL{ Phys. Rev. Lett.} 
 
\def\PRE{{ Phys. Rev.} E}

\usepackage{graphicx}
\usepackage{color}
\usepackage{dcolumn}
\usepackage{bm}
\usepackage{amsmath}

\begin{document}
\title{ The role of topology in electrical properties of bacteriorhodopsin and rat olfactory receptor I7}

\author{E. Alfinito}
\email {eleonora.alfinito@unisalento.it}
\author{L. Reggiani}

\affiliation{Dipartimento di Ingegneria dell'Innovazione, 
Universit\`a del Salento, Via Monteroni, 73100 Lecce, Italy (EU); CNISM - Consorzio Nazionale Interuniversitario per le Scienze Fisiche della Materia, via della Vasca Navale 84, 00146 Roma, Italy (EU)  
}

\date{\today}

\begin{abstract}
We report on electrical properties of the two sensing proteins: bacteriorhodopsin and rat olfactory receptor OR-I7.
As relevant transport parameters we consider the small-signal 
impedance spectrum and the static current-voltage characteristics. 
Calculations are compared with available experimental results and the model predictability is tested for future perspectives. 
\end{abstract}

\maketitle

%
Sensing proteins are of relevant interest because of the fundamental role they play in living environments at a cellular level \cite{heptahelices}.
Recently, much work has been devoted to the investigation of the 
 G protein-coupled receptors (GPCRs), a large family of proteins sensitive to the capture of single or a few specific molecules (ligands) and a similar protein, the bacteriorhodopsin (bR), a light sensitive protein whose action is related to a proton pump.
The activation of sensing proteins is a quite complex mechanism which in any case starts with a conformational change.  
The capture mechanism and also their 3D (tertiary) structure  is quite the same for all of them, despite of the very different nature of the possible ligands (from photons to neurotoxins, from smells to hormones). 
The attention on these receptors is very high and the attempts of characterization cover different protein features.
On this respect, we mention recent experiments on the current-voltage (I-V) characterization of bR nanolayers when sandwiched between metallic contacts \cite{Jin,Jin1}, and when using an atomic force microscope technique \cite{Gomila}. 
On the same subject,  recent experiments \cite{Hou} were performed 
on the rat OR-I7 anchored on a functionalized gold substrate with the technique of molecular self-assembly. 
These measurements \cite{Hou,Hou1} showed the possibility to monitor the protein sensing action, and the related conformational change, by means of the modification of the electrochemical impedance spectrum (EIS) in the presence of a controlled flux of specific odorants (heptanal, octanal).
\par
To our knowledge, the above experimental results did not frame into
a unified theoretical scheme, neither from a macroscopic point of view
\cite{Wang}, nor from a microscopic point of view \cite{tao06,migliore09}.
As an alternative to these approaches we developed a coarse grained model which connects electrical and morphological protein modifications by using an impedance network protein analogue (IPNA) \cite{Nano,Life,Jap,Epl}.
Here, the model is further validated on available experiments concerning the small signal impedance spectrum of rat OR-I7 and its predictability is tested by a comparative investigation on bR and rat OR-I7.  
\par
\begin{figure}
\centering
\resizebox{0.35\textwidth}{!}{%
\includegraphics{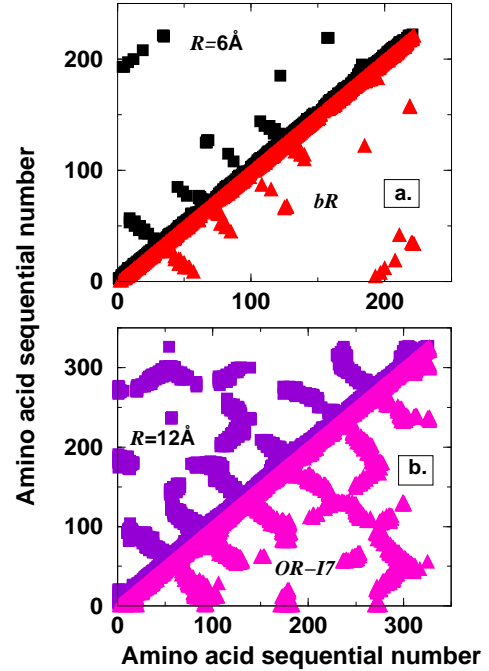}
}
\caption{(Color online) Contact map of the native and activated state of bacteriorhodopsin (fig.1a) and OR-I7 (fig.2b). Fig.1a: the native state (black full circles) vs.  the activated state (red full triangles) for \eR= 6 \aa.
Fig.1b: the native state (violet full squares) vs.  the activated state (magenta full triangles) for \eR= 12 \aa.
%
}
	\label{fig:fig_1}
\end{figure}
To set up the IPNA, the protein tertiary structure, in the native and activated state, have to be known. This information is available in the protein data banks,  although for few proteins  \cite{PDB}.  
When these data have been gained, the protein in a given state can be mapped into a topological network (graph) \cite{Tirion,lenovere09}, each node of the graph corresponding to a $C_{\alpha}$, taken as the single interacting center. 
When two $C_{\alpha}$s are closer than an assigned cut-off distance, \eR, a link is drawn between the corresponding nodes \cite{Jap}. 
Thus, by fixing the value of \eR, we fix the complexity of the graph. 
For  bR we take the couple of PDB entries 2NTU (native) and 2NTW (activated) \cite{PDB}. 
For rat OR-I7 we used the GPCR bovine rhodopsin as a template and the MODELLER software to construct, by homology, the tertiary structure for both the native and activated states \cite{Jap}.
\par
A comparative analysis between the native and activated
states of bR and rat OR I7, is given by their contact maps \cite{Albert}
as reported in Figs. 1a and 1b, respectively.
For a fixed cut-off radius, these maps provide a reduced representation  of the main differences between the configurations, as one can deduce by comparing the left (native-state) and right (activated-state) part with respect to the diagonal in Figs. 1a and 1b.
Here, we have found that the most significative contrast between the native and activated state is obtained for  \eR =6 \aa in bR and for  \eR =12 \aa in rat OR-I7. 
\par
The graph representation is then used to describe the protein physical features.
The graph turns into the INPA, by attributing an elemental impedance $Z_{i,j}=l_{i,j}/    
\mathcal{A}_{i,j}(\rho^{-1} + \rm{i} \epsilon_{i,j}\, \epsilon_0\omega)$ \cite{Nano, Life},  to each link. 
%
%
\begin{figure}
\centering
\resizebox{0.35\textwidth}{!}{%
\includegraphics{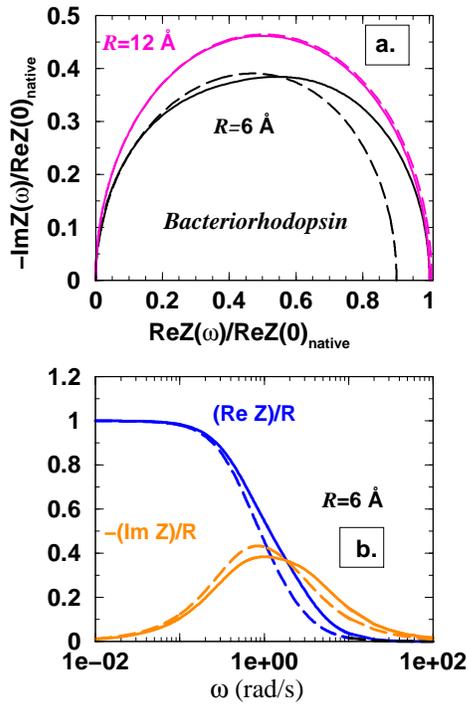}
}
\caption{(Color online) Fig2a: Nyquist plot of bR for the PDB entries 2NTU, the native state, (continuous line)  and 2NTW, the activated state, (dashed line). 
Plots are calculated for \eR = 6 \aa (black)  and 12 \aa (magenta), and values are normalized to the zero-frequency value of the impedance pertaining to the native state.
Fig2b: Normalized real (blue) and imaginary (orange) parts of the total impedance of bR versus the angular frequency. The data are calculated for an interaction radius \eR= 6\aa. Continuous lines refer to the native state, dashed lines to the activated state.}
	\label{fig:fig_2}
\end{figure}
The elemental impedance is constituted by a an Ohmic resistance connected in parallel with a parallel plate capacitor, filled with a dielectric, so that  both charge motion and electrical polarization can be described. 
Finally, the model was implemented to determine non-linear  I-V  characteristics by adopting a sequential tunneling mechanism of charge transfer, as detailed in Refs. (\onlinecite{Epl}).
We assume a tunneling probability as given by the WKB approximation: 
$P_{i,j}= \exp[- \frac{2 l_{i,j}}{\hbar} \sqrt{2m(\Phi-eV_{i,j})} ] $,
%
where $V_{i,j}$ is the  potential drop between the  $i$-th and $j$-th node,
$m$ is an effective electron mass, here taken as that of the free electron, and $\Phi$ is the barrier height, taken, for these calculations, equal to
59 meV.  
%
%
%
%
\begin{figure}
\centering
\resizebox{0.35\textwidth}{!}{%
\includegraphics{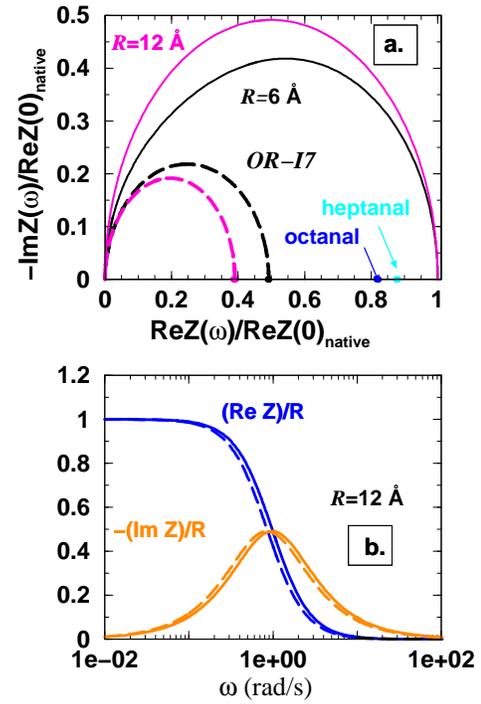}
}
\caption{(Color online) Fig2a: Nyquist plot of rat OR-I7, in the native state, (continuous line) and the activated state, (dashed line). 
Plots are calculated for \eR = 6 \aa (black)  and 12 \aa (magenta), and values are normalized to the zero-frequency value of the impedance pertaining to the native state.
Fig2b: Normalized real (blue) and imaginary (orange) parts of the total impedance of rat OR-I7 versus the angular frequency. The data are calculated for an interaction radius \eR= 12\aa. Continuous lines refer to the native state, dashed lines to the activated state.}
	\label{fig:fig_3}
\end{figure}
\par 
Figure 2a reports the Nyquist plots  for the native and activated state of bR and for  \eR=6 and 12 \aa, respectively. 
The plots exhibit an almost semicircular shape, slightly suppressed because of the presence of many relaxation-time scales \cite{Barsoukov}. 
The contrast between the two states takes the highest level with \eR=6 \aa,  lowering  systematically for increasing  cut-off values.  
In particular, the static impedance is found to decrease for about 10\%, at most, when moving from the native to the activated state. 
To complete the analysis of the impedance spectrum,
figure 2b reports the frequency dependence of the real, $Re\{Z(\omega)\}$, and imaginary, $Im\{Z(\omega)\}$, parts of the impedance (Bode plots), respectively, for the highest contrast of \eR=6 \aa. 
To our knowledge, experimental data on bR are  not  available.   
%
%
\par
Figure 3a
reports the Nyquist plot for the native and activated state 
of rat OR-I7  and for  \eR=6 and 12 \aa, respectively. 
In analogy with the case of bR, the plots exhibit an almost semicircular shape. 
The contrast between the two states takes the highest level with 
\eR=12 \aa.  
In particular,  the static impedance is found to decrease for about 
60\%, at most, when moving from the native to the activated state. 
Both the shape of the plot and the magnitude of the resistance change  are in qualitative agreement  with EIS experiments \cite{Hou1} obtained for a self assembled layer of OR-I7  in the presence of its specific odorants heptanal and octanal. 
In Fig. 3a,  the two points under the arrows on the x axes report
the dose response of OR-I7 layered structure to a concentration of 10$^{-4}$ mol/L of octanal and heptanal.
In analogy with the case of bR, Fig. 3b reports the normalized real and imaginary parts of Z (Bode plots)  as function of $\omega$,  for the highest contrast of \eR = 12 \aa, respectively.  
We would remark the difference between the Nyquist plots of bR and OR-I7,
the former showing the maximal resolution of the native and activated configurations for \eR = 6 \aa, the latter for \eR =12 \aa. 
The origin of this difference is attributed to  a more massive conformational change in the olfactory receptor than in the bR, as already documented by the contact maps in Fig. 1a and 1b. 
%
\begin{figure}
\centering
\resizebox{0.45\textwidth}{!}{%
\includegraphics{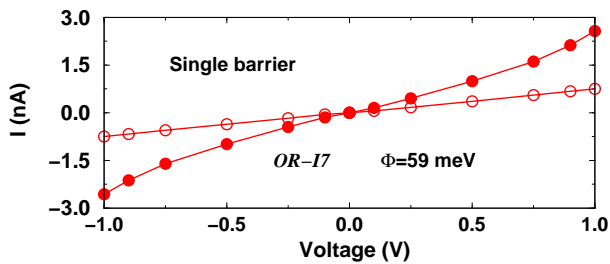}
}
\caption{(Color online) I-V characteristics of the native and activated states of rat OR-I7.
Empty circles refer to the native state, and full circles to the activated state, respectively. 
The continuous lines are guides for eyes. }
	\label{fig:fig_4}
\end{figure}
\par
Figure 4 reports the I-V characteristics for the rat OR-I7, obtained with \eR=6 \aa and the same transport parameters used for bR \cite{Epl}. 
With respect to  bR,  we have found a larger difference between the current response of the native and activated state, with the native
state showing a quasi-linear response in the range of considered voltages.
By contrast, the current corresponding to the activated
state exhibits a significant super-linear increase, in close analogy with what observed in  bR experiments. 
This result can be interpreted in terms of large distances between amino-acids in the native state, which oppose the tunnelling mechanism of charge transfer, and of a significant decreasing of the same distances after the conformational change, which favors charge transfer. 
This interpretation applies well also to the large difference exhibited by the shapes of the Nyquist plots for the OR-I7, and confirms the strict correlation between the protein tertiary structure and its conformational change.
\par
Finally, we stress that present theoretical results take into account only the topological structure and its modifications, and refer to a protein completely activated.
Furthermore,  different bio-chemical contributions, as the role of the environment, are  neglected. 
Indeed, being independent of configurational change, the environment should play a minor role in determining the change of electrical properties induced by the conformational change, which justifies the omission of its effects.
\par
%
In conclusion, we have investigated the correlation between a conformational change and the variation of electrical transport properties in a sensing protein. 
%
%
Theory  makes use of an appropriate procedure devised to account for the interaction responsible for charge transfer between neighbouring amino-acids.  
On this basis, we have investigated the change of the small-signal impedance spectrum and that of the static I-V characteristics in bacteriorhodopsin and rat OR-I7. 
Concerning the small-signal impedance spectrum, the theory is compared with  EIS experiments carried out on self-assembled monolayers of OR-I7, an odour sensitive GPCR protein, over functionalized gold substrates.
Qualitative agreement with experiments is found when considering  a maximum interaction radius between amino acids of 12 \aa,  which evidences a substantial modification of the Nyquist plot (decrease of the resistance up to a maximum of about 60$\%$) in the presence of specific odorants, like heptanal and octanal.
Concerning the I-V characteristics, the theory applied to a nanolayer of OR-I7 evidences a substantial increase of the current, at a given voltage, when going from the native to the activated state. 
By testing the predictability of the model,  
for the Nyquist plot of bR we found a small but still significant deviation of the spectrum with a decrease of the resistance for a maximum of about 10 $\%$ when going from the native to the activated state.
\vskip1pc\noindent  
This research is supported by the European Commission under the Bioelectronic Olfactory Neuron Device (BOND) project within the grant agreement number 228685-2.

\begin{thebibliography}{99}

\bibitem{heptahelices}
R. J. Lefkowitz, \Journal{Nature Cell. Bio.}{2}{E133} {2000};
%
L. Buck, R. Axel, 
\Journal{Cell}{65}{175}{1991}.

%
\bibitem{Jin}
Y. D. Jin, N. Friedman, M. Sheves, T. He, and D. Cahen, \Journal{PNAS}{103} {8601} {2006}.
%
\bibitem{Jin1}
Y.D. Jin, N. Friedman, M. Sheves, and D. Cahen, \Journal{Adv. Funct. Mater.}{17}{1417} {2007}.
%
\bibitem{Gomila}
I. Casuso, L. Fumagalli, J. Samitier, E. Padr{\'o}s, L. Reggiani, V. Akimov, G. Gomila, \Journal{Nanotechnology}{18}{2007}{465503};
I. Casuso, L. Fumagalli, J. Samitier, E. Padr{\'o}s, L. Reggiani, V. Akimov, G. Gomila, \Journal{\PRE} {76}{041919 }{2007}.
%
\bibitem{Hou} 
Y. Hou,   S. Helali, A. Zhang, N. Jaffrezic-Renault, C. Martelet, 
J. Minic, T. Gorojankina, M.-A. Persuy, E. Pajot-Augy, R. Salesse,  F. Bessueille, J. Samitier, A. Errachid, V. Akimov, L. Reggiani, C. Pennetta, E. Alfinito,  \Journal{Biosensors \& Bioelectronics} {21}{1393} {2006}.
%
\bibitem{Hou1}
Y. Hou,  N. Jaffrezic-Renault, C. Martelet, A. Zhang, J. Minic, T. Gorojankina, M.-A. Persuy, E. Pajot-Augy, R. Salesse, V. Akimov, L. Reggiani, C. Pennetta, E. Alfinito, O. Ruiz, G.l Gomila, J. Samitier, A. Errachid, \Journal{Biosensors \& Bioelectronics} {22}{1550} {2007}.
%
\bibitem{Wang}
W. Wang, T. Lee and W. A. Reed, \Journal{Rep. Prog. Phys.}{68}{523}{2005}.
%
\bibitem{tao06}
N.  Tao, \Journal{Nature Nanotechnology}{1}{173}{2006}.
%
\bibitem{migliore09}
A. Migliore, S. Corni, D. Varsano,  M. Klein, and R. Di Felice,  
\Journal{\JPCB}{113}{9402}{2009}.
%

\bibitem{Nano} 
E. Alfinito, C. Pennetta and L. Reggiani, \Journal{Nanotechnology} {19} {065202}{ 2008}.
%
\bibitem{Life}
C. Pennetta,  V. Akimov, E. Alfinito, L. Reggiani, T. Gorojankina, J. Minic, E. Pajot, M.A. Persuy, R. Salesse,  I. Casuso, A. Errachid, G. Gomila, O. Ruiz,  J. Samitier, Y. Hou,  N. Jaffrezic-Renault, G. Ferrari, L. Fumagalli and M. Sampietro, \textit{Nanotechnology of the Life Science} ed.
C.S.S.R. Kumar (Wiley-VCH, Weinheim 2006) vol 4, p. 217.
%
\bibitem{Jap}
E. Alfinito, C. Pennetta and L. Reggiani,\Journal{\JAP}{105}{084703} {2009}, and references therein. 
%
\bibitem{Epl}
E. Alfinito and L. Reggiani, \Journal{\EPL}{105}{68002} {2009}.
%
\bibitem{PDB}
H. M. Berman, J. Westbrook, Z. Feng, G. Gilliland, T.N. Bhat, H. Weissing, I.N. Shindyalov and P. Bourne, \Journal{Nucleic Acids Research}{28}{235}{2000}.
%
 %

%
\bibitem{Tirion}
M. M. Tirion, \Journal{\PRL}{77}{1996}{1905}; 
C. Micheletti, P. Carloni and A. Maritan, \Journal{Proteins}{55}{635}{2004}.
%
\bibitem{lenovere09}
N. Le Nov\`ere, M. Hucka, H. Mi, S. Moodie, F. Schreiber, A. Sorokin, E. Demir, K. Wegner, M.I. Aladjem, S.M. Wimalaratne, F.T. Bergman, R. Gauges, P. Ghazal, H. Kawaji, L. Li, Y. Matsuoka1, A. Vill\'eger, S.E. Boyd, L. Calzone, M. Courtot, U. Dogrusoz, T.C. Freeman, A. Funahashi, S. Ghosh, A. Jouraku, S. Kim, F. Kolpakov, A. Luna, S. Sahle, E. Schmidt, S. Watterson, G. Wu, I. Goryanin, D.B. Kell, C. Sander, H. Sauro, J.L. Snoep, K. Kohn and H. Kitano, \Journal{Nature Biotechnology}{27}{735}{2009}.
%
%
\bibitem{Albert}
R. Albert, and A.~L. Barabasi,
\Journal{Rev. Mod. Phys.}{74}{47}{2002}
%
%
\bibitem{Barsoukov}
E. Barsoukov and J. Ross Macdonald, \textit{Impedance Spectroscopy: Theory, Experiment, and Applications} (John Wiley and Sons, Inc., Hoboken, New Jersey
2005)
%
\end{thebibliography}
\end{document}